\documentclass[12pt,a4paper,final]{iopart}

\usepackage{iopams}  
\usepackage{graphicx}
\usepackage{subfigure}
\usepackage{color}

\begin{document}

\title[Transient dynamics of a colloidal particle driven through a viscoelastic fluid]{Transient dynamics of a colloidal particle driven through a viscoelastic fluid}

\author{Juan Ruben Gomez-Solano$^{1,2}$ and Clemens Bechinger$^{1,2}$}
\address{$^{1}$2. Physikalisches Institut, Universit\"at Stuttgart, Pfaffenwaldring 57, 70569 Stuttgart, Germany}
\address{$^{2}$Max-Planck-Institute for Intelligent Systems, Heisenbergstrasse 3, 70569 Stuttgart, Germany}

\eads{\mailto{r.gomez@physik.uni-stuttgart.de}}

\begin{abstract}
We experimentally study the transient motion of a colloidal particle actively dragged by an optical trap through different viscoelastic fluids (wormlike micelles, polymer solutions, and entangled $\lambda$-phage DNA). We observe that, after sudden removal of the moving trap, the particle recoils due to the recovery of the deformed fluid microstructure. We find  that the transient dynamics of the particle proceeds via a double exponential relaxation, whose relaxation times remain independent of the initial particle velocity whereas their amplitudes strongly depend on it. 
While the fastest relaxation mirrors the viscous damping of the particle by the solvent, the slow relaxation results from the recovery of the strained viscoelastic matrix.
We show that this transient information, which has no counterpart in Newtonian fluids, can be exploited to investigate linear and nonlinear rheological properties of the embedding fluid, thus providing a novel method to perform transient rheology at the micron-scale.

\end{abstract}

\vspace{2pc}

\section{Introduction}

Viscoelasticity is ubiquitous in many materials ranging from biological fluids, polymers, micellar systems, colloidal suspensions and more. In general, such materials are characterized by a strongly time-dependent mechanical response to stress or strain. For instance, in case of an oscillatory shear, they may exhibit either liquid or solid-like properties depending on the imposed frequency~\cite{larson}.  Such non-Newtonian behavior originates from the storage and dissipation of energy within their complex microstructure which gives rise to a finite macroscopic stress-relaxation time~\cite{ashwin}. Given the abundance of systems which involve the interplay between microscale viscoelastic flows and embedded micron-sized objects, as typically found in lab-on-a-chip devices~\cite{kim,romeo,lee}, porous media~\cite{scholz,howe}, and in active matter~\cite{lauga,martinez,qin}, the understanding of the equilibrium and  non-equilibrium dynamics of colloidal particles in viscoelastic fluids is a noteworthy topic in soft matter. Microrheology has become in recent years a field which successfully addresses some of these problems.
In particular, it provides alternative methods to bulk rheology to investigate flow and deformation properties of microlitre samples of viscoelastic materials. For example, linear shear moduli can be inferred in passive microrheology from the thermal fluctuations of suspended colloidal particles via a generalized Stokes-Einstein relation~\cite{gittes,buchanan,larsen,gomezsolano}. On the other hand, in active microrheology small-amplitude oscillatory forces can be applied to the particle, which enables the dynamical measurement of the linear viscoelastic response of the surrounding fluid~\cite{mizuno,wilhelm,sriram0,gomezsolano1}. More recent developments of active microrheology have aimed to probe nonlinear response of complex fluids by perturbing their microstructure far away from equilibrium. This has been achieved by dragging or rotating the probe through the fluid at sufficiently large constant force or torque \cite{cappallo,wilking,rich,choi,cribb}, or at constant velocity~\cite{khan,sriram,chapman,gomezsolano2}, thus inducing e.g. thinning.

Despite multiple potential applications of microrheology, so far most of the experiments have focused on the study of linear-response or steady-state quantities. More complex transient behavior observed macroscopically, e.g. nonlinear creep, stress relaxation and strain recovery, is far less well understood within the context of microrheology. Indeed, theoretical works suggest that the drag force acting on a colloidal probe driven through a non-Newtonian fluid, strongly depends on the  spatio-temporal deformation of the microstructure of the surrounding fluid ~\cite{squires,depuit,zia}. Therefore, in contrast to Newtonian liquids, where a low Reynolds number guarantees simple Stokesian flows and linear-response drag forces, it is questionable whether this is applicable when a particle is dragged by a time-dependent driving force over large distances through a viscoelastic medium~\cite{swan}. Then, it is not evident which kind of rheological properties of the fluid can be extracted by simply tracking the position of a probe in response to a pulsed or Heaviside time-dependent driving force. It should be mentioned that such a time-dependent force represents the microrheological analogue to a macroscopic flow startup/cessation experiment~\cite{lopezbarron}, where the time evolution of the stress (strain)  is measured upon suddenly imposing or relieving strain (stress), respectively. Although relevant for further developments of microrheology and for the general understanding of the motion of micron-sized particles in viscoelastic media, these questions have not been experimentally investigated so far.

In this work, we study the local microstructural recovery of several viscoelastic fluids by means of microrheology. For this purpose, we drag spherical particles at constant velocity through them using optical tweezers, thus inducing strain, and then we studdenly remove the trapping force. We observe that the particle recoils after removal of the trap until the complete recovery of the investigated fluid. We find that the transient particle dynamics during the recovery proceeds via a double exponential relaxation, whose relaxation times remain independent of the initial particle velocity whereas their amplitudes strongly depend on it. 
These processes reveal how the elastic energy stored by the fluid due to a large local strain is eventually dissipated. While the fastest relaxation mirrors the viscous damping of the particle by the solvent, the slow process results from the relaxation of the viscoelastic matrix.
In particular, we show that  this experimental method allows to probe linear and nonlinear transient rheological behavior of the viscoelastic fluid under study and to extract unambiguously its stress-relaxation time. In addition, our findings have important consequences on the understanding and interpretation of the motion of colloids actively driven through viscoelastic media. Thus, we provide clear experimental evidence that, unlike passive and small-amplitude oscillatory microrheology, the active motion of a colloidal particle over large distances does not only probe the bulk viscoelastic properties of the fluid but is also affected by the flow field and deformation imposed by the particle.

\section{\label{sec:exp}Experimental description}

\begin{figure}
	\center
	\includegraphics[width=0.7\columnwidth]{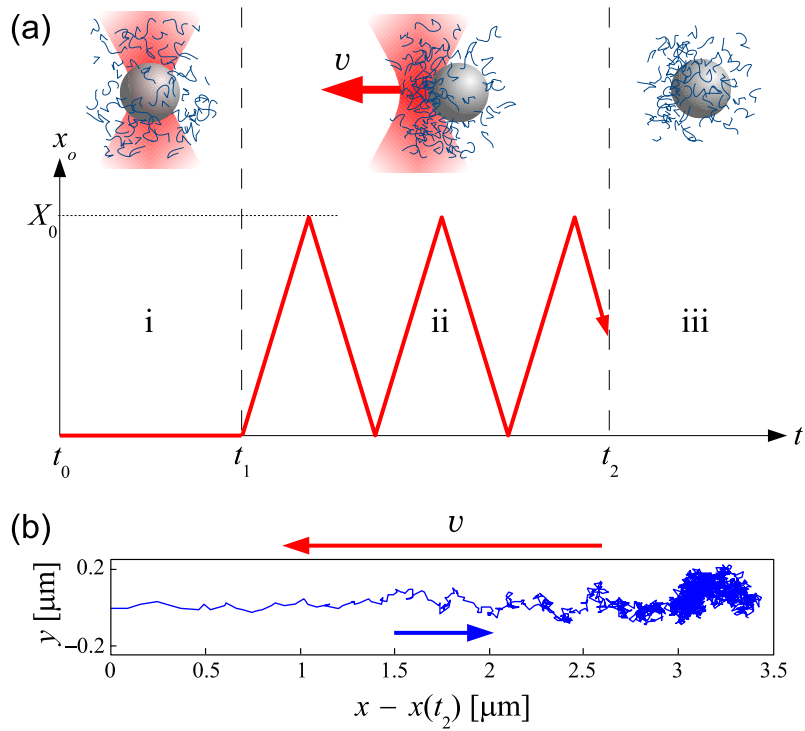}
	\caption{(a) Experimental protocol to study the transient motion of a colloidal probe during the recovery of a viscoelastic fluid:  (i) passive, (ii) active, and (iii) transient microrheology. (b) Trajectory of a $r = 1.6\,\mu\mathrm{m}$ particle suspended in a wormlike micellar solution (CPyCl/NaSal 7mM, $T = 303$~K) during the  microstructural recovery after removal of the trap ($k = 8\,\mathrm{pN}\,\mu\mathrm{m}^{-1}$, $v = 3.20\,\mu\mathrm{m}\,\mathrm{s}^{-1}$). The blue arrow indicates the direction of the particle recoil. See text for explanation.}
\label{fig:fig1}
\end{figure}

In our experiments, we prepare equimolar solutions of cetylpyridinium  chloride (CPyCl) and sodium salicylate (NaSal) in deionized water at concentrations between 5 and 9 mM. We also studied samples of other viscoleastic fluids, namely, aqueous polyacrylamide (PAAm) solutions ($M_w=18 \times 10^{6}$ at 0.05  \% wt) and entangled $\lambda $-phage DNA solutions (500 $\mu$g/mL in 10 mM Tris-HCl and 1 mM EDTA). 
We add to each viscoelastic solution a small volume fraction of spherical silica particles, whose radii are chosen between $r = 0.9$ and 2.2~$\mu$m. The temperature $T$  of each sample is kept fixed  by a flow thermostat  between $20\pm0.1$ and $40\pm 0.1^{\circ}$C. Then, a single particle is trapped by optical tweezers created by deflection of a Gaussian laser beam ($\lambda = 1070$~nm) on a galvanostatically driven pair of mirrors and subsequent focusing by a microscope objective ($100\times$, NA = 1.3) into the sample. In order to avoid hydrodynamic interactions, the particle is trapped at least $20\,\mu\mathrm{m}$ away from the walls of the sample cell. The optical trap at position ${\bf{r}}_o = (x_o, y_o)$ exerts a restoring force on the particle at position ${\bf{r}} = (x,y)$, i.e. ${\bf{f}}_o=-k ({\bf{r}} - {\bf{r}}_o) $. 
Using video microscopy, we track the center of mass of the particle at 120 frames~$\mathrm{s}^{-1}$ with a spatial accuracy of 10~nm.
We implemented the protocol sketched in Fig.~\ref{fig:fig1}(a), where the red solid line represents the time evolution of $x_o$, to study the recovery of the different viscoelastic fluids. 
\begin{itemize}

	\item[(i)]{During time $t_0 \le t \le t_1$, the trap is kept at rest $(x_o = 0, y_o=0)$, so that the particle is in thermal equilibrium with the surrounding fluid.  The thermal  fluctuations of $(x,y)$ enable us to determine the stiffness $k$ of the optical trap and to characterize the linear microrheological properties of the fluids by passive microrheology.}

	\item[(ii)]{Then, at time $t = t_1$, we start to move the trap at constant velocity $v$ along a linear path, $(x_o =v(t-t_1)$, $y_o = 0)$, thus actively dragging the particle through the fluid along $x$ and inducing strain. At position $x_o = X_0$, where $X_0 = 16 \, \mu\mathrm{m} \gg r$, the motion of the trap is reversed: $v \rightarrow -v$. We perform a sequence of these back-and-forth movements of the trap position until the distance travelled by the particle during a period $\tau_o =  2X_0 / v$ remains constant. 
Depending on the value of the driving period $\tau_o$ we can induce either linear or nonlinear microrheological response of the viscoelastic fluid~\cite{gomezsolano2}. 
During this step, we measure the effective viscosity of the fluid $\eta = f_d/(6\pi r v)$ by active microrheology from the balance between the drag force $f_d$ and  $|{\bf{f}}_o|$.}

	\item[(iii)]{Finally, at time $t = t_2$ and position $x_o=X_0/2$ the optical trap is suddenly turned off.
During this final step, the particle motion at time $t>t_2$  is subjected to the relaxation of the previously strained microstructure of the fluid.
The subsequent recovery results in a recoil of the particle along the $x$-direction opposite to that of the prior trap motion. This recoil is illustrated in  Fig.~\ref{fig:fig1}(b), where we plot an example of a 2D trajectory of the particle during the recovery process of a wormlike micellar solution after removal of the moving trap. We observe that, in absence of external forces, the particle gradually performs Brownian motion once the fluid completely equilibrates.}

\end{itemize}
Note that, in a Newtonian fluid, the particle undergoes Brownian diffusive motion immediately after removal of the force and therefore such a recoil is not observed, as verified in our experiments in $\mathrm{H}_2\mathrm{O}$ and glycerol (data not shown). 
Since the recoil takes place in the direction parallel to the driving force, in the following we only concentrate on the time evolution of its $x$ coordinate. 
We define $t_2 =0\, \mathrm{s}$ as the time at which the trap is turned off and $\Delta x(t) = x(t) - x(t_2)$ as the distance travelled backward by the particle during the recovery of the fluid. For a fixed velocity $v$ of the trap, the previous protocol is repeated at least 20 times in order to perform an ensemble average of $\Delta x$. We typically span values of $v$ between $0.4\,\mu\mathrm{m}\, \mathrm{s}^{-1}$ and $40\,\mu\mathrm{m}\, \mathrm{s}^{-1}$. 
For these values of $v$ the Deborah number is between 0.025 and 2.5, whereas the Weissenberg number is between 0.125 and 12.5, thus fully probing both linear and nonlinear viscoelasticity of the investigated fluids.

\section{\label{sec:results}Results}

\begin{figure}
	\center
	\includegraphics[width=0.8\columnwidth]{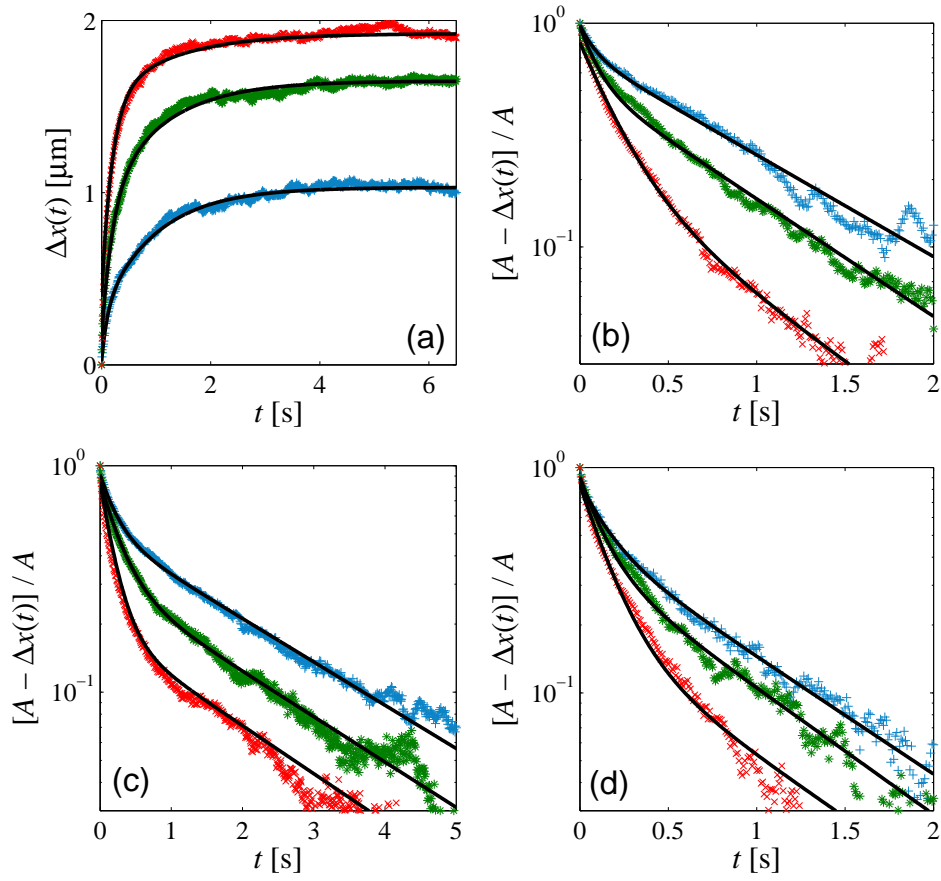}
\caption{(a) Time evolution of particle displacement $\Delta x(t)$ during the recovery of a wormlike micellar solution after removing the trap, initially moving at different velocities $v$. From bottom to top: $v = 1.60 \,\mu\mathrm{m}\,\mathrm{s}^{-1}$ ($+$),  $6.40 \,\mu\mathrm{m}\,\mathrm{s}^{-1}$ ($*$), $32.00 \,\mu\mathrm{m}\,\mathrm{s}^{-1}$ ($\times$). Time evolution of the normalized particle displacement $[A - \Delta x(t)]/A$ for (b) a wormlike micellar solution, same data and symbols as in Fig.~\ref{fig:fig2}(a), (c) PAAm solution at $v = 2.13 \,\mu\mathrm{m}\,\mathrm{s}^{-1}$ ($+$), $4.00 \,\mu\mathrm{m}\,\mathrm{s}^{-1}$ ($*$), $12.80 \,\mu\mathrm{m}\,\mathrm{s}^{-1}$ ($\times$), and (d) $\lambda-$DNA at $v = 1.30 \,\mu\mathrm{m}\,\mathrm{s}^{-1}$ ($+$), $2.10 \,\mu\mathrm{m}\,\mathrm{s}^{-1}$ ($*$), $3.20 \,\mu\mathrm{m}\,\mathrm{s}^{-1}$ ($\times$). The solid lines are fits to Eq.~(\ref{eq:eq0}).}
\label{fig:fig2}
\end{figure}

In Fig.~\ref{fig:fig2}(a) we plot the time evolution of the displacement $\Delta x(t)$ of a $r=1.6\,\mu\mathrm{m}$ particle after removal of the trap moving at different initial velocities $v$ in a wormlike micellar solution (CPyCl/NaSal 5 mM, $T = 303.16$~K). We find that $\Delta x(t)$ strongly depends on $v$, with a monotonic increase over a finite time-scale until the complete relaxation of the system to thermal equilibrium, at which $\Delta x(t)$ saturates to a constant value $A \equiv \Delta x(t \rightarrow \infty)$.  In order to understand this dynamics, in Fig.~\ref{fig:fig2}(b)  we represent in a semi-log plot the time evolution of the normalized displacement, ${[A -\Delta x(t)]}/{A}$. As a result all curves start at 1 and decay monotonically to 0 regardless of the value of $v$. We observe that all the curves exhibit an initial fast decay, followed by a slower one, which suggests the existence of two exponential relaxations. Indeed, we verify that for all $v$ the time evolution of $\Delta x(t)$ can be fitted to the function
\begin{equation}\label{eq:eq0}
	\Delta x(t) = A - A_{\mathrm{f}} \exp\left( - \frac{t}{\tau_{\mathrm{f}}}\right) - A_{\mathrm s} \exp\left( - \frac{t}{\tau_{\mathrm{s}}}\right),
\end{equation}
with two relaxation times $\tau_{\mathrm{f}}$ and $\tau_{\mathrm{s}}$  and two amplitudes $A_{\mathrm{f}}$ and $A_{\mathrm{s}}$ which satisfy $A = A_{\mathrm s} + A_{\mathrm f}$. This is shown in Figs.~\ref{fig:fig2}(a) and ~\ref{fig:fig2}(b), where the fits to Eq.~(\ref{eq:eq0}) are represented as solid lines. Furthermore, we find that this transient behavior of the particle position is not specific to the recovery of wormlike micellar solutions. For example, in Figs.~\ref{fig:fig2}(c) and~\ref{fig:fig2}(d) we plot the time evolution of ${[A -\Delta x(t)]}/{A}$ for a $r=1.6\,\mu\mathrm{m}$ particle recoiling in a PAAm solution and in $\lambda$-phage DNA, respectively, for different $v$. We find that, in these cases  Eq.~(\ref{eq:eq0}) also describes the time evolution of  $\Delta x(t)$, even when the rheological properties of such viscoelastic fluids are more complex than those of wormlike micelles~\cite{zhu}.

\begin{figure}
	\includegraphics[width=\columnwidth]{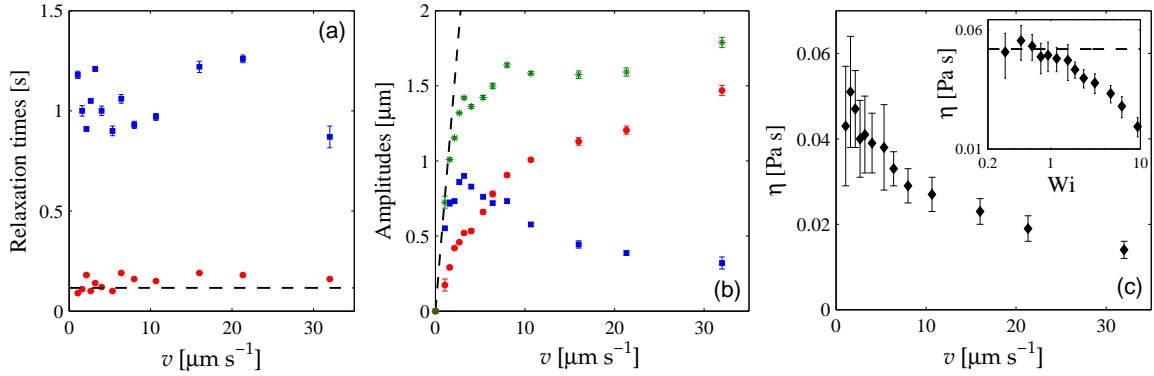}
\caption{Dependence of (a) relaxation times $\tau_{\mathrm{s}}$ ($\square$), $\tau_{\mathrm{f}}$ ($\circ$), and b) amplitudes $A_{\mathrm{s}}$ ($\square$), $A_{\mathrm{f}}$ ($\circ$), obtained from the fit to Eq.~(\ref{eq:eq0}), on the driving velocity $v$. The symbols ($*$) correspond to $A=A_{\mathrm{s}}+A_{\mathrm{f}}$, whereas the dashed lines represent the values $\tau_{\mathrm{eff}}$ and $A_{\mathrm{eff}}$ computed from Eq.~(\ref{eq:eq4}) for the  rheological model~(\ref{eq:eq1}). (c) Effective viscosity $\eta$ of the fluid as a function of the driving velocity $v$ determined by active microrheology during step (ii). Inset: log-log plot of $\eta$ as a function of the corresponding Weissenberg number, same data as in main Figure. The dashed line represents the value of the zero shear viscosity $\eta_0$.}
\label{fig:fig3}
\end{figure}

To support the existence of two distinct relaxation processes, we studied the dependence of the fitting parameters of Eq.~(\ref{eq:eq0}) on the velocity $v$. For instance, in Fig.~\ref{fig:fig3}(a) we plot the two relaxation times $\tau_{\mathrm{f}}$ and $\tau_{\mathrm{s}}$ as a function of $v$ during the recovery of a wormlike micellar solution (CPyCl/NaSal 5 mM, same sample as in Fig.~\ref{fig:fig2}(a)). We find that the values of these two time-scales are different by one order of magnitude, and interestingly, both are independent of $v$. For this reason, we will refer to them in the following as \emph{fast} and \emph{slow} relaxation times, respectively. Their corresponding amplitudes,  $A_{\mathrm{f}}$ and $A_{\mathrm{s}}$, are plotted in  Fig.~\ref{fig:fig3}(b). We observe that, unlike the relaxation times, the amplitudes strongly depend on $v$. In particular, while $A_{\mathrm{f}}$ increases monotonically with increasing $v$ and then it levels off, $A_{\mathrm{s}}$ has a maximum and then decreases with increasing $v$. 

\begin{figure}
	\center
	\includegraphics[width=\columnwidth]{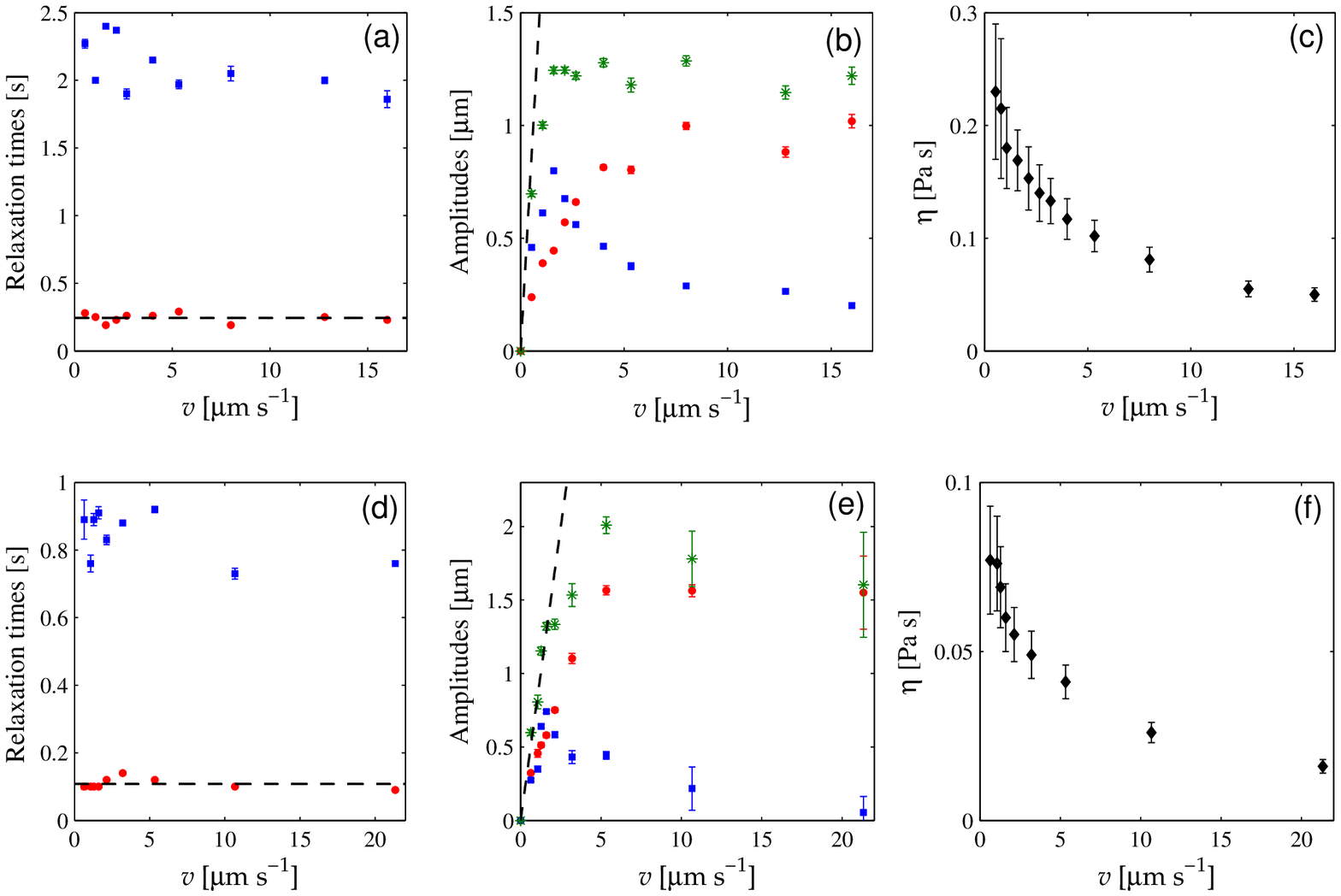}
\caption{Dependence on imposed velocity $v$ of (a) relaxation times $\tau_{\mathrm{s}}$ ($\square$), $\tau_{\mathrm{f}}$ ($\circ$), (b) amplitudes $A_{\mathrm{s}}$ ($\square$), $A_{\mathrm{f}}$ ($\circ$), $A=A_{\mathrm{s}}+A_{\mathrm{f}}$ ($*$) and (c) effective viscosity $\eta$ ($\diamond$) for a $r = 1.6\,\mu{\mathrm{m}}$ particle in a polymer solution of PAAm.  Dependence on $v$ of (d) relaxation times $\tau_{\mathrm{s}}$ ($\square$), $\tau_{\mathrm{f}}$ ($\circ$), (e) amplitudes $A_{\mathrm{s}}$ ($\square$), $A_{\mathrm{f}}$ ($\circ$), and $A=A_{\mathrm{s}}+A_{\mathrm{f}}$ ($*$) and (f) effective viscosity $\eta$ ($\diamond$) for a $r = 1.6\,\mu{\mathrm{m}}$ particle in $\lambda-$DNA. The dashed lines in Figs.~\ref{fig:figPAAMDNA}(a) and \ref{fig:figPAAMDNA}(d) correspond to $\tau_{\mathrm{eff}}$ whereas in Figs.~\ref{fig:figPAAMDNA}(b) and \ref{fig:figPAAMDNA}(e) they represent $A_{\mathrm{eff}}$. See text for explanation.}
\label{fig:figPAAMDNA}
\end{figure}

For the values of $v$ we span in our experiments, both linear and nonlinear microrheological response of the viscoelastic fluid is induced by the particle motion previously driven by the tweezers. This is shown in Fig.~\ref{fig:fig3}(c)  where we plot the effective viscosity $\eta$ of the fluid, measured by active microrheology during step (ii) of Fig.~\ref{fig:fig1}(a), as a function of the imposed velocity $v$. As highlighted in the inset of Fig.~\ref{fig:fig3}(c), at small $v \lesssim 3 \,\mu\mathrm{m}\,\mathrm{s}^{-1}$, $\eta$ is constant and equal to the zero-shear viscosity $\eta_0 = 0.045\pm 0.005$~Pa~s (dashed line), because the strain rate imposed by the particle is sufficiently slow for the fluid to lose its memory on previous microstructural deformations. 
Consequently, Newtonian-like behavior is observed through the validity  of the linear-response Stokes law: $f_d = 6\pi r \eta_0 v$. On the other hand, at sufficiently large $v \gtrsim 3 \,\mu\mathrm{m}\,\mathrm{s}^{-1}$, the particle motion is able to induce an orientational order of the fluid microstructure. As a result, the fluid exhibits thinning, where $\eta$ decreases dramatically with increasing $v$~\cite{gomezsolano2}. 
Note that the emergence of nonlinear rheological effects of the fluid  due to the local deformation created by the particle are better quantified by means of the Weissenberg number Wi. In our system, Wi can be estimated by the product of the characteristic rate of deformation $\frac{\mathrm{d}}{\mathrm{d}t}\left( \frac{x_o(t)}{2r} \right) = \frac{v}{2r}$ times the largest relaxation time of the system $\tau_{\mathrm{s}}$: $\mathrm{Wi} = v \tau_{\mathrm{s}}/(2r)$. Then, nonlinear non-Newtonian behavior is expected at $\mathrm{Wi} \gtrsim 1$, at which the local microstructural deformation of the fluid is so strong that the Stokes law must break down. Indeed, in Fig.~\ref{fig:fig3}(c) we show that for the spanned values of $v$, the effective viscosity of the fluid exhibits the aforementioned transition at $\mathrm{Wi} \approx 1$ and therefore we cover both linear and nonlinear microrheological behavior.
Hence, our results demonstrate that the double-exponential relaxation of the particle position is a rather general process during the recovery of viscoelastic fluids with a well-defined zero-shear viscosity, regardless of the value of the Weissenberg number.

We point out that the dependence on the velocity $v$  of the relaxation times $\tau_{\mathrm{f}}$ and $\tau_{\mathrm{s}}$, and the amplitudes $A_{\mathrm{f}}$ and $A_{\mathrm{s}}$ is not particular to the particle motion during the recovery of wormlike micelles, but also found for the other investigated viscoelastic fluids.  In Figs.~\ref{fig:figPAAMDNA}(a)-(c) we plot the results for an aqueous semidilute polymer solution of PAAm  ($M_w=18 \times 10^{6}$ at 0.05~\%~wt, temperature $T=293.16$~K), whereas the results for $\lambda$-phage DNA (500 $\mu$g/mL in 10 mM Tris-HCl and 1 mM EDTA, temperature $T=298.16$~K) are displayed in Fig.~\ref{fig:figPAAMDNA}(d)-(f). Once again, we observe that the two distinct relaxation time-scales remain independent of $v$ both in the linear and nonlinear response regime, whereas the amplitudes exhibit a very similar dependence to that observed for the wormlike micellar solution. Thus, our findings suggest that only the amplitudes of $\Delta x(t)$ and not its relaxation times encode rheological information of the local deformation of the fluid microstructure created by the particle, the latter defining the time-scales at which the system retains memory on the energy stored by such an elastic deformation.

\section{\label{sec:disc}Discussion}

\begin{figure}
\begin{subfigure}
  \centering
  \includegraphics[width=.475\linewidth]{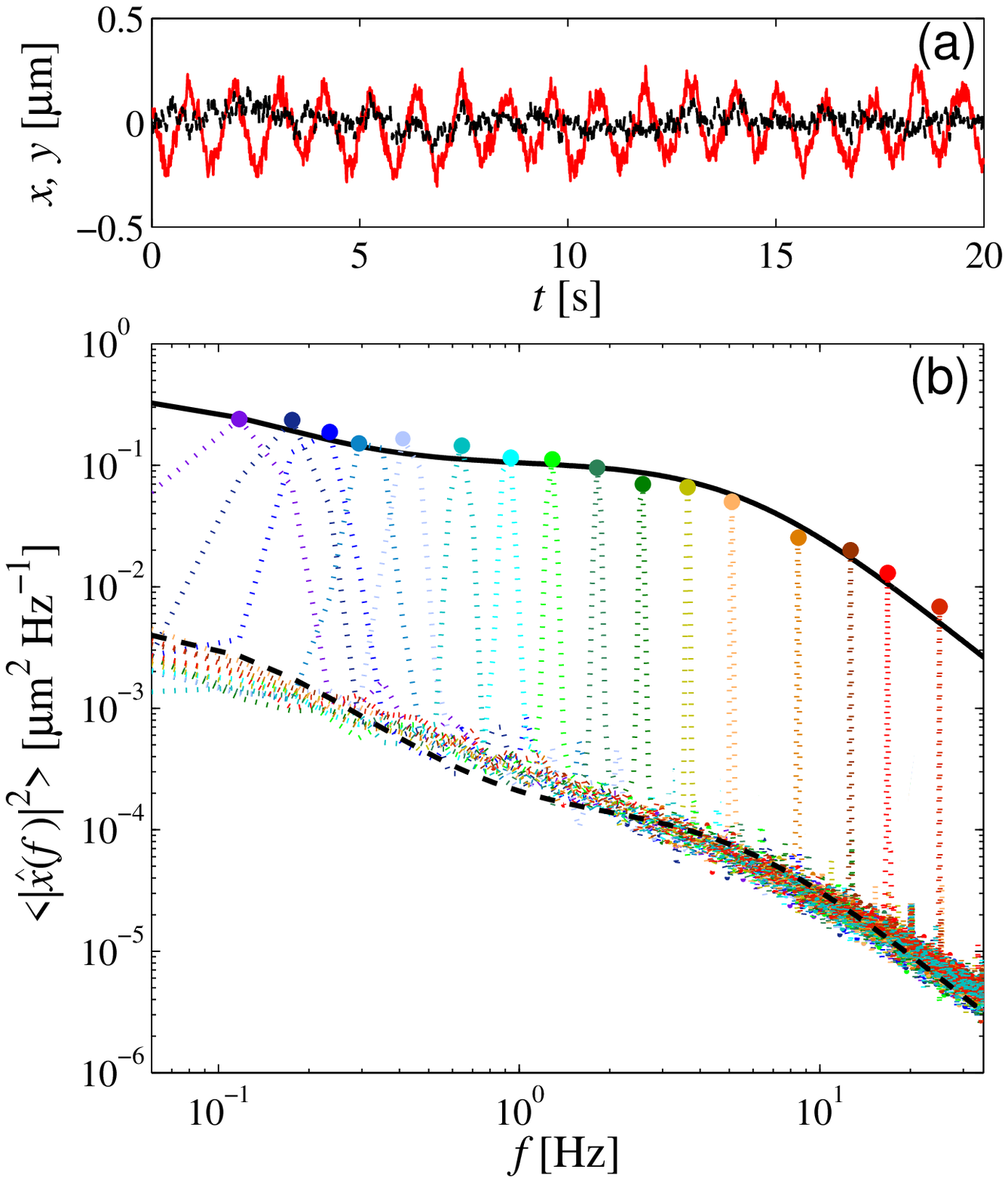}
\end{subfigure}%
\begin{subfigure}
  \centering
  \includegraphics[width=.525\linewidth]{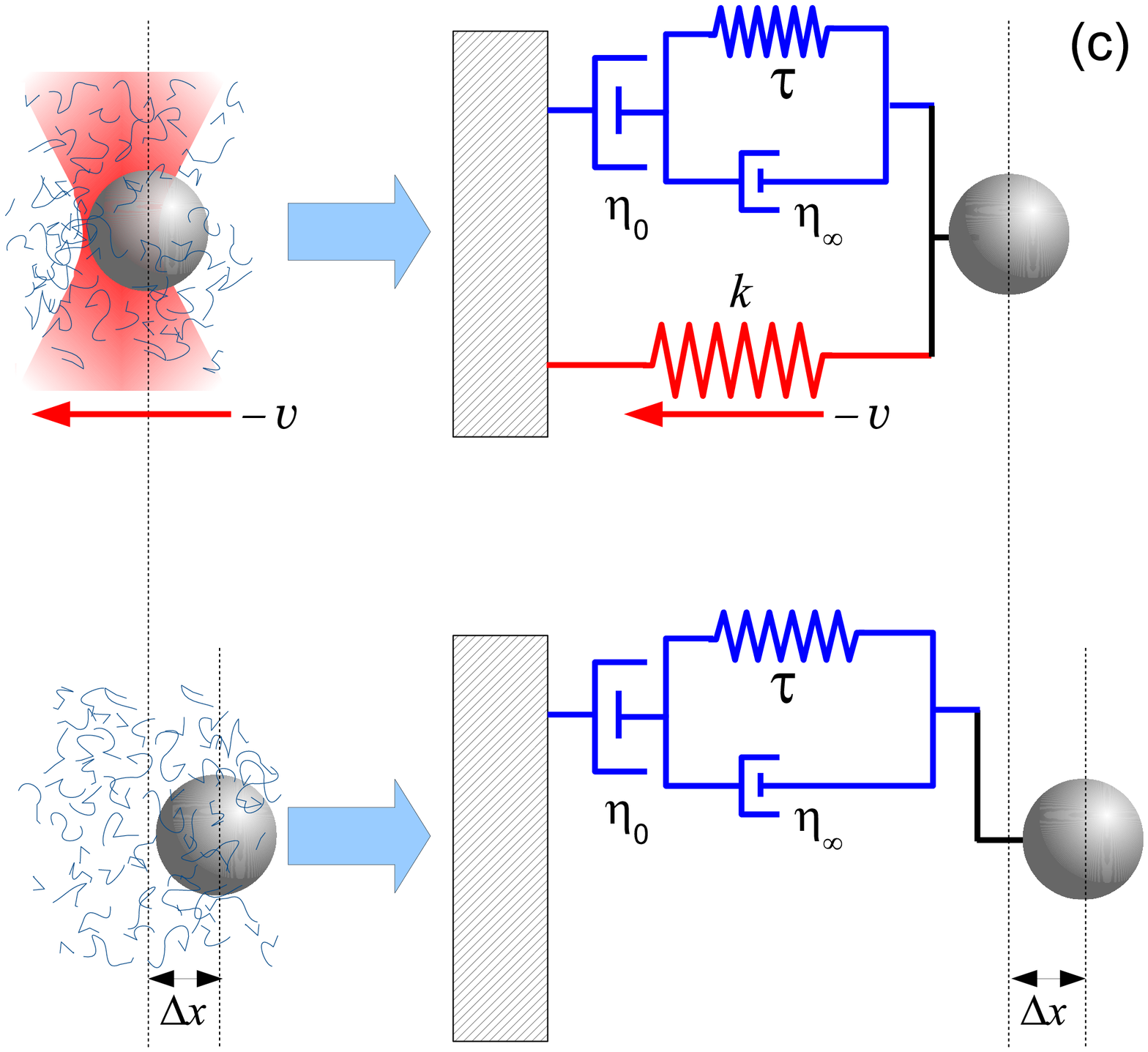}
\end{subfigure}
\caption{(a) Time evolution of the coordinates $x$ (red solid line) and $y$ (black dashed line) of a $r = 1.6\,\mu\mathrm{m}$ particle in a wormlike micellar solution in response to a small-amplitude sinusoidal motion of the optical trap ($k = 2\times 10^{-6}\,\mathrm{N}\,\mathrm{m}^{-1}$) with amplitude $X_0 = 200$~nm and angular frequency $\Omega = 2\pi$~rad~s$^{-1}$. (b) Experimental power spectral density of $x$ (dotted lines) at different oscillation frequencies $\Omega/(2\pi)$ of the trap. The peaks at each imposed $\Omega/(2\pi)$  are indicated by circles. The dashed line represents Eq.~(\ref{eq:PSDpas}), whereas the solid line corresponds to the total power spectral density described by Eq.~(\ref{eq:PSDx}). (c) Schematic representation of the particle dynamics dragged at constant velocity  $-v$ through a fluid with linear shear modulus~(\ref{eq:eq1}) (upper panel) and during the relaxation after removing the driving force (lower panel). See text for explanation.}
\label{fig:figjeffreys}
\end{figure}

In order to gain an understading of this obviously generic behavior of $\Delta x(t)$, we now investigate whether a Langevin model can reproduce the particle dynamics after sudden removal of the driving force after dragging the particle over a large distance $X_0 > 2r$ through the fluid, i.e. after inducing a large strain $X_0/(2r)$. For this purpose we consider the simplest case of a particle driven by the moving trap in a wormlike micellar solution. 
In general, under sufficiently small applied shear strain rate $\dot{\gamma}$, the resulting shear stress of a viscoelastic fluid $\sigma$ is determined  by the linear shear modulus $G$: $\sigma(t) = \int_{-\infty}^t G(t - t') \dot{\gamma}(t')\,\mathrm{d}t'$. In the case of a wormlike micellar solution, $G$ can be simply characterized by a single stress relaxation time $\tau$, the viscosity of the solvent $\eta_{\infty}$ and the zero-shear viscosity $\eta_0$ (Jeffreys model)~\cite{raikher}
\begin{equation}\label{eq:eq1}
	G(t) = 2\eta_{\infty} \delta(t) + \frac{\eta_0 - \eta_{\infty}}{\tau} \exp\left(-\frac{t}{\tau}\right).
\end{equation}
Therefore, according to commonly accepted assumptions, the steady-state dynamics of the $x$ coordinate of the particle is described in the linear-response regime by the non-Markovian Langevin equation~\cite{fodor}
\begin{equation}\label{eq:eq2}
	\int_{-\infty}^t \Gamma(t - t') \dot{x}(t') \,\mathrm{d}t' = - k[x(t) - x_o(t)] + \xi(t).
\end{equation}
In Eq.~(\ref{eq:eq2}), $\Gamma(t - t') = 6\pi r G(t - t')$ is a memory kernel related to the shear modulus $G$ of the fluid given in Eq.~(\ref{eq:eq1}), whereas $\xi$ is a coloured Gaussian noise with autocorrelation function $\langle \xi(t) \xi(t')\rangle = k_BT\Gamma(|t - t'|)$. 

\subsection{Transient motion under small deformation}

We first experimentally verify that  the Langevin Eq.~(\ref{eq:eq2}) correctly describes the equilibrium dynamics of the particle and its linear response to small-amplitude oscillatory forces exerted by the optical trap. This implies that, for sufficiently small deformations of the fluid, bulk rheological properties can be unambiguously determined by measuring the position of the embedded particle, as widely assumed in passive and linear active microrheology. For this purpose, we apply to the particle a sinusoidal force $kx_o(t)$ by moving the trap according to $x_o(t) = X_0\sin(\Omega t)$, with amplitude $X_0 = 200\,\mathrm{nm} \ll 2r$ at different driving frequencies $\Omega/(2\pi)$. The characteristic strain of the fluid locally induced by this oscillatory motion is $X_0 / (2r) = 0.06$. The typical time evolution of the particle position in response to such a driving force is illustrated in  Fig.~\ref{fig:figjeffreys}(a). We observe that, while the coordinate $x$ responds to the force at the same excitation frequency $\Omega/(2\pi)$, the coordinate $y$ remains unaffected. Therefore, linear response dynamics can be tested by means of $x$, whereas thermal equilibrium properties can be analysed by means of $y$. This can be done, e.g. by comparing the experimental power spectral densities of such stochastic signals to those predicted by the Langevin model ~(\ref{eq:eq2}). For a sinusoidal driving force, Eq.~(\ref{eq:eq2}) yields the following expressions for the power spectral densities of $x$ and $y$ in terms of the linear storage $G'(f)$ and loss $G''(f)$ modulus of the fluid
\begin{equation}\label{eq:PSDx}
	\langle | \hat{x}(f) |^2 \rangle = S_{pas}(f) + S_{act}(f),
\end{equation}
\begin{equation}\label{eq:PSDy}
	\langle | \hat{y}(f) |^2 \rangle = S_{pas}(f),
\end{equation}
where
\begin{equation}\label{eq:PSDpas}
	S_{pas}(f)  = \frac{2 k_B T}{\pi f}\frac{6\pi r G''(f)}{[k + 6\pi rG'(f)]^2 + [6\pi r G''(f)]^2},
\end{equation}
\begin{equation}\label{eq:PSDact}
	S_{act}(f) = \frac{k^2 X_0^2}{2}\frac{\delta(f - \frac{\Omega}{2\pi})}{[k + 6\pi rG'(f)]^2 + [6\pi r G''(f)]^2},
\end{equation}
are the contributions to the power spectral density due to thermal fluctuations and to the active driving force, respectively. In particular, the prefactor $k_B T$ in Eq.~(\ref{eq:PSDpas}) results from the thermal origin of the fluctuations, while the delta function $\delta(f - \frac{\Omega}{2\pi})$ in Eq.~(\ref{eq:PSDact}) mirrors the mechanical response of the system at frequency $f = \frac{\Omega}{2\pi}$.
Indeed, we find the expressions (\ref{eq:PSDx})-(\ref{eq:PSDact}) derived from Eqs.~(\ref{eq:eq1})~and~(\ref{eq:eq2}) correctly describe the measured power spectral densities of $x$ and $y$ for all the investigated frequencies. This in turn validates such a Langevin approach in the linear response regime, provided that the particle performs small displacements in the fluid, i.e. when sufficiently small flow and deformations are induced. We show this in Fig.~\ref{fig:figjeffreys}(b), where we demonstrate that, at the unexcited frequencies  $f \neq \frac{\Omega}{2\pi}$, the power spectral density is correctly described Eq.~(\ref{eq:PSDpas}) (dashed line), where $G'(f)$  and $G''(f)$ are obtained from the Fourier transform of Eq.~(\ref{eq:eq1}). On the other hand, an excellent agreement between the experimental peaks of $\langle | \hat{x}(f) |^2 \rangle$  at the excited frequencies $f = \frac{\Omega}{2\pi}$ (circles) and Eq.~(\ref{eq:PSDx}) (dashed line) is observed.

Once we have verified that the Langevin model (\ref{eq:eq2}) correctly describes the particle dynamics under sufficiently small deformations of the fluid, we now focus on its transient behavior after removal of the driving force. A schematic representation of the average forces acting on the particle is shown in Fig.~\ref{fig:figjeffreys}(c).
Interestingly, Eq.~(\ref{eq:eq2}) predicts that after dragging the particle at velocity $-v$ at time $t<0$ through a medium with linear viscoelasticity (\ref{eq:eq1}), it recoils with an initial velocity $(\eta_0 - \eta_{\infty})v / \eta_{\infty}$ at time $t=0$ upon removal of the trap. Moreover, the subsequent displacement $\Delta x(t)$ evolves in time $t > 0$  according to a single exponential relaxation
\begin{equation}\label{eq:eq3}
	\Delta x(t) = A_{\mathrm{eff}}\left[  1 - \exp\left( - \frac{t}{\tau_{\mathrm{eff}}}\right) \right],
\end{equation}
with amplitude and relaxation time 
\begin{equation}\label{eq:eq4}
	A_{\mathrm{eff}} = \left(1  -  \frac{\eta_{\infty}}{\eta_0} \right) v\tau, \,\,\, \tau_{\mathrm{eff}} = \frac{\eta_{\infty}}{\eta_0} \tau, 
\end{equation}
respectively. 
See~\ref{appa} for a more detailed derivation of Eqs.~(\ref{eq:eq3}) and (\ref{eq:eq4}).
We notice that, according to Eq.~(\ref{eq:eq4}), the amplitude $A_{\mathrm{eff}}$ depends linearly on $v$ and encodes information of the rheological properties of the fluid. Indeed, the driven motion of the particle induces a strain in the fluid, which is captured in the linear rheological model of Eq.~(\ref{eq:eq1}) as an effective elastic compression $\sim v \tau$, as depicted in the upper panel of Fig.~\ref{fig:figjeffreys}(c). This results in a total displacement of the particle proportional to $v$ upon removing the driving force, as schematized in the lower panel of Fig.~\ref{fig:figjeffreys}(c). 
On the other hand, the relaxation time $\tau_{\mathrm{eff}}$ of the particle displacement is independent of $v$ and always smaller than the stress-relaxation time of the fluid $\tau$. This is due to presence of the solvent, whose viscosity $\eta_{\infty}$ makes the particle become at rest in the effective Langevin model (\ref{eq:eq2}) before the fluid completely relaxes its stress initially stored by the elastic compression induced during the active driving process.

\subsection{Transient motion under large deformation}
We now compare the values of the effective parameters  $\tau_{\mathrm{eff}}$ and $A_{\mathrm{eff}}$ with those we obtain experimentally for the colloidal particle after performing large displacements $X_0 > 2r$ through the fluid, i.e. after inducing a characteristic strain $X_0/(2r) = 5$.
In Fig.~\ref{fig:fig3}(a) we show  as a dashed line the value of the effective relaxation time $\tau_{\mathrm{eff}}$, where the parameters $\eta_0$, $\eta_{\infty}$ and $\tau$ are determined by passive microrheology during step (i) of Fig.~\ref{fig:fig1}(a). We observe that the values of the fast relaxation time $\tau_{\mathrm{f}}$ obtained from the fit to Eq.~(\ref{eq:eq0}) (circles) agree well with $\tau_{\mathrm{eff}}$, which implies that the fast relaxation process mirrors the effective elastic relaxation of the probe damped by the solvent viscosity. Nevertheless,  in Fig.~\ref{fig:fig3}(b) we show that $A_{\mathrm{eff}} >  A_{\mathrm{f}}$ for all $v>0$ which suggests that the fast relaxation is just one of the possible channels for the complete relaxation of the system to thermal equilibrium. As a matter of fact, in Fig.~\ref{fig:fig3}(b) we show that at small $v$, $A = A_{\mathrm s} + A_{\mathrm f}$ agrees well with $A_{\mathrm{eff}}$. Thus, $A$ actually quantifies the elastic compression of the fluid induced by the driven particle in the linear response regime. Systematic deviations of $A$ from $A_{\mathrm{eff}}$ show up with increasing $v$, though, because nonlinear microrheological response of the fluid is induced during the previous driven motion of the particle. In such a case, Eq.~(\ref{eq:eq1}) fails to describe the viscoelastic properties of the fluid. We confirm this in Fig.~\ref{fig:fig3}(c), where we show that at small $v \lesssim 3 \,\mu\mathrm{m}\,\mathrm{s}^{-1}$, i.e. $\mathrm{Wi} \lesssim 1$, linear microrheological response is induced, i.e. $\eta = \eta_0$, whereas  at sufficiently large $v \gtrsim 3 \,\mu\mathrm{m}\,\mathrm{s}^{-1}$, i.e. $\mathrm{Wi} \gtrsim 1$, the fluid microstructure responds nonlinearly and the system exhibits thinning. Note that the values of $v$ for which thinning emerges agree well with those for which large deviations of $A$ from $A_{\mathrm{eff}}$ are observed. Nonlinear microrheological behavior of the fluid affects separately the behavior of $A_{\mathrm f}$ and $A_{\mathrm s}$ as well. In particular, $A_{\mathrm f}$ exhibits a saturation whereas $A_{\mathrm s}$ has a maximum and then decreases for the values of $v$ at which prior thinning is observed. These results clearly demonstrate that nonlinear response of a viscoelastic fluid, induced by active microrheology, gives rise to nonlinear transient behavior of the probe upon removal of the active driving.

We also check that the previous mechanism applies to the recovery of the polymer solution and of $\lambda-$DNA. In absence of simple rheological models for the linear shear modulus $G(t)$ of such viscoelastic fluids, in analogy to the observations for the wormlike micellar solutions, we first assume that $\tau_{\mathrm{eff}} = \tau_{\mathrm{f}}$ and $\tau = \tau_{\mathrm{s}}$, which implies that $\eta_{\infty} = \eta_0 \tau_{\mathrm{eff}}/\tau = \eta_0 \tau_{\mathrm{f}}/\tau_{\mathrm{s}}$.
Then, accordingly to  Eq.~(\ref{eq:eq4}) we compute the effective amplitude by means of 
\begin{equation}\label{eq:eqaff}
	A_{\mathrm{eff}} =   v (\tau_{\mathrm{s}} - \tau_{\mathrm{f}}).
\end{equation}
Surprisingly, we observe in Figs.~\ref{fig:figPAAMDNA}(b) and~\ref{fig:figPAAMDNA}(e) that the values of $A_{\mathrm{eff}}$ obtained from Eq.~(\ref{eq:eqaff}) (dashed lines) agree very well with the total amplitude $A=A_{\mathrm{s}}+A_{\mathrm{f}}$ (asterisks) at sufficiently small velocities $v$  for both fluids, thus making evident the robustness of the two distinct time-scales $\tau_{\mathrm{f}}$ and $\tau_{\mathrm{s}}$ regardless of the details of the rheological properties of the embedding viscoelastic fluid.
On the other hand, for velocities at which thinning is induced, i.e. at which $\eta$ decreases with increasing particle velocity, the amplitudes exhibit a dependence very similar to that observed for wormlike micelles. While $A_{\mathrm{f}}$ saturates with increasing $v$, $A_{\mathrm{s}}$ has a maximum and then decreases. The consequence is that the total amplitude $A$ levels off with increasing velocity, thus signalling the emergence of a nonlinear regime where the elastic compression of the fluid induced by the particle becomes independent of $v$. Hence, our results demonstrate the  generality of the transient microrhelogical behavior for very distinct viscoelastic fluids both in the linear and nonlinear response regimes.

\begin{figure}
	\center
	\includegraphics[width=0.95\columnwidth]{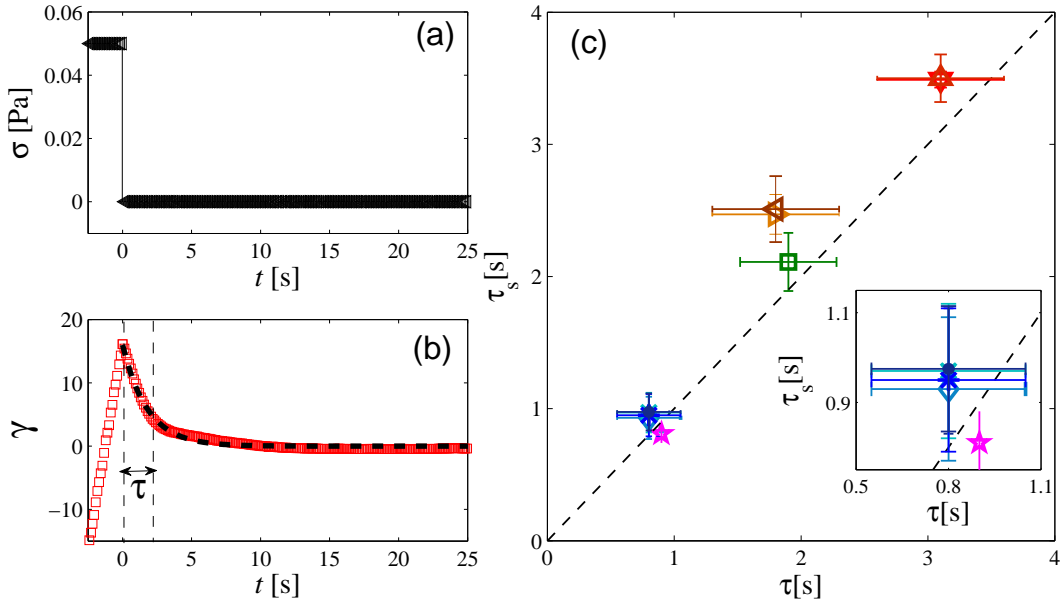}
\caption{(a) Example of time evolution of imposed stress and (b) resulting strain during a shear-flow cessation experiment of a micellar solution (CPyCl/NaSal 7~mM) in a Couette geometry. The dashed line in Fig.~{\ref{fig:fig4}}(b) represents an exponential fit with relaxation time $\tau = 1.8~\mathrm{s}$.
(c) Comparison between the slow relaxation time $\tau_{\mathrm{s}}$ of various viscoelastic fluids determined by the transient motion of particles of different radii $r$ and the equilibrium stress-relaxation time $\tau$. The symbols are: CPyCl/NaSal 5~mM, $r=0.9\,\mu$m ($\times$), $r=1.3\,\mu$m ($\diamond$), $r=1.6\,\mu$m ($\ast$), $r=2.2\,\mu$m ($\bullet$); $\lambda-$phage DNA, $r=1.6\,\mu$m ($\star$); PAAm, $r=1.6\,\mu$m ($\Box$);  CPyCl/NaSal 7~mM, $r=1.3\,\mu$m ($\triangleleft$),  $r=2.2\,\mu$m ($\triangleright$); CPyCl/NaSal 9~mM, $r=1.3\,\mu$m ($\triangledown$),  $r=1.6\,\mu$m ($\vartriangle$). Inset: expanded view of main figure around 0.9~s.}
\label{fig:fig4}
\end{figure}

The previous findings make evident that the transient rheological behavior of viscoelastic fluids probed by microrheology does not only depend on the characteristic rate of local deformation $v/(2r)$, but also on the total strain $X_0/(2r)$. On one hand, for small $X_0/(2r) \ll 1$, the Langevin model~(\ref{eq:eq2}) correctly describes the dynamics of the particle either in thermal equilibrium with the surrounding fluid or externally driven into linear response regime by small-amplitude forces, where vanishingly small deformations of the fluid are induced.
On the other hand, a different transient behavior is observed for sufficiently large strain $X_0/(2r) \gtrsim 1$, where such a  Langevin model ceases to be valid.
In particular,  it fails to reproduce the slow relaxation inferred from Eq.~(\ref{eq:eq1}) after cessation of a large deformation imposed by the driven particle. The reason is that Eq.~(\ref{eq:eq2}) only accounts for the coarse-grained drag force exerted by the fluid. Such a viscoelastic drag force is enough to describe equilibrium thermal fluctuations of the particle position without recourse to any information on the surrounding flow field. However, it does not incorporate the effect of the transient flow and deformation localized around the particle when dragged over large distances, and which cannot be neglected during the recovery process of the fluid. In the simplistic Langevin picture, the only possible channel for the system to reach thermal equilibrium is by means of the effective relaxation of the particle position. While this relaxation probes the linear bulk properties of the fluid, it ignores the complete stress relaxation of the local viscoelastic flow field. 

In fact, the aforementioned drawbacks suggest that, in reality, the residual slow relaxation represents a second mechanism to completely release the stress stored by the strained fluid. The two relaxation mechanisms allow in turn the fluid to reach thermal equilibrium in conjunction with the particle motion. These observations hint at the stress-relaxation time $\tau$ of the fluid as the slow relaxation time $\tau_{\mathrm{s}}$. 
In order to test this hypothesis, we perform using a rotational rheometer (HAAKE RheoStress 1, double-gap cylinder geometry) the macroscopic version of the transient microrheology experiments, This is illustrated in Figs.~\ref{fig:fig4}(a) and \ref{fig:fig4}(b), where we plot an example of the time evolution of the shear stress $\sigma$ applied to a wormlike micellar solution, and the resulting shear strain $\gamma$, respectively. After reaching a steady state with constant shear rate $\dot{\gamma}$ at time $t < 0$, the shear stress is suddently removed at $t = 0$, and afterwards $\gamma$ decays in time, until the fluid completely recovers. Then, we extract the relaxation time $\tau$ of the fluid from the exponential fit $\gamma(t) = \gamma_0 \exp(-t/\tau)$.
In Fig.~\ref{fig:fig4}(c) we plot $\tau_{\mathrm{s}}$, as a function of the corresponding $\tau$ for all the investigated viscoelastic fluids. We find that within the experimental errors, both time-scales are equal. In Fig.~\ref{fig:fig4}(c) we also verify that the relaxation times $\tau_{\mathrm{s}}$ are independent of the size of the probe particle. See~\ref{appb} for more detailed results obtained for different particle sizes.
The independence of $\tau_{\mathrm{s}}$ from $v$, $r$ and $X_0$, i.e. from the detailed flow and strain field around the particle, demonstrates that it is a physical property inherent to the viscoelastic fluid. Thus, this time-scale fixes a value above which the memory on the storage of elastic energy by the fluid microstructure is lost.
Hence, the the slow transient dynamics of a colloidal probe after removal of a driving force represents a new straightforward method to measure the stress-relaxation time of viscoelastic fluids, provided that the previous driven motion induces a sufficiently large strain. This can be directly carried out by means of particle tracking without the need of any model for the dynamics or any further elaborate Fourier or Laplace analysis, as commonly done in passive and small-amplitude active microrheology~\cite{wilson, tassieri}.

We point out that the slow relaxation is most prominent close to the onset of nonlinear microrheological response around $\mathrm{Wi} \approx 1$, as verified by the presence of the maxima of $A_{\mathrm{s}}$ in Figs.~\ref{fig:fig3}(b), \ref{fig:figPAAMDNA}(b) and \ref{fig:figPAAMDNA}(e). 
Therefore, our findings have important consequences on the interpretation of transient behavior in active microrheology under large deformations~\cite{wilking,chapman,brasovs}. This is because they provide experimental evidence that, even at small Weissenberg number, where only linear microrheological response is expected to occur, the response of the colloidal particle is strongly affected by the relaxation of the local flow and strain field around it. This is in contrast to passive and active microrheology under small-amplitude deformations, where only bulk properties of the fluid are probed by the particle. In addition, our results have significant implications for the behavior of micron-sized objects in transient viscoelastic flows, such as those typically found in microchannels, where large strains are commonly induced by high flow rates and sharp constrictions. In such systems, unlike the case of Newtonian liquids, the conventional Langevin description for the suspended particles involving only coarse-grained drag forces and noise must break down. Furthermore, our results are particularly important for the understanding and modelling of the motion of biological microswimmers, e.g. bacteria and spermatozoa, whose natural environment is frequently viscoelastic~\cite{lauga,martinez,qin}. Such active systems usually operate at high Weissenberg number and travel large distances through the viscoelastic medium, thus inducing complex flow fields and large strains. Therefore, the relaxation of the elastic energy stored by the surrounding fluid microstructure can strongly impact their swimming mechanisms through the different processes found in our experiments. 

\section{Summary and conclusion}

We have experimentally studied the motion of a colloidal probe during the local strain recovery of several viscoelastic fluids upon shutoff of an active driving force. 
We have shown that, even when a Langevin model correctly describes linear response dynamics of the particle position under vanishingly small strain, it cannot capture its transient dynamics during the recovery process when large deformations are previously induced by large displacements of the particle through the fluid.
In particular, we have found that this transient dynamics proceeds via a double exponential decay, which reveals the cooperative relaxation to thermal equilibrium of the particle and the surrounding strained fluid. We have shown that the fastest relaxation mirrors the viscous damping of the particle by the solvent, whereas the slow one results from the relaxation of the viscoelastic matrix. 
Thus, the two relaxation processes represent dissipation channels for the elastic energy stored by the fluid microstructure upon cessation of a sufficiently large deformation. We have demonstrated that, while the amplitudes of these relaxation modes encode either linear or nonlinear microrheological information of the fluid induced during the prior driving, their relaxation times are insensitive to it. Consequently, this transient microrheological method allows to determine unambiguously stress-relaxation times of micro-litre samples of viscoelastic fluids.

\section*{Acknowledgments}
We thank C. Lozano for helpful discussions. We acknowledge financial support of the Deutsche Forschungsgemeinschaft, BE 1788/10-1.

\appendix

\section{Transient particle's dynamics in a Jeffreys fluid}\label{appa}
For the Langevin model of a particle moving in a Jeffreys fluid, i.e. Eqs.~(\ref{eq:eq1}) and (\ref{eq:eq2}), an analytical expression for the time evolution of its position $\Delta x(t)$ after cessation of an externally applied force can be readily derived. This can be achieved by realizing that, for $\eta_0 > \eta_{\infty} > 0$ and $\tau > 0$, the non-Markovian equation of motion (\ref{eq:eq2}) for $x$ can be written as a linear system of two Markovian Langevin equations~\cite{villamaina}
\begin{eqnarray}\label{eq:app1}
  	\dot{x}(t) & = & \frac{1}{\tau}\left(\frac{\eta_0}{\eta_{\infty} }- 1\right )[u(t) - x(t)] - \frac{k[ x(t) - x_o(t)]}{6\pi\eta_{\infty}r} + \sqrt{\frac{k_BT}{3 \pi \eta_{\infty}r}} \zeta_1(t),\nonumber \\
	\dot{u}(t)& = & \frac{1}{\tau} [x(t) -  u(t)] + \sqrt{\frac{k_BT}{3\pi(\eta_0 - \eta_{\infty})r}} \zeta_2(t),
\end{eqnarray}
where $\zeta_1$ and $\zeta_2$ are white noises of mean $\langle \zeta_i \rangle = 0$ and correlation $\langle \zeta_i(t) \zeta_j(t') \rangle = \delta_{i,j} \delta(t-t')$, $i,j=1,2$, whereas the auxiliary variable $u$ is defined as
\begin{equation}\label{eq:appu}
	u(t) = \frac{1}{\tau} \int_{-\infty}^t \exp\left(  -\frac{t - t'}{\tau} \right) \left[  x(t') + \tau \sqrt{\frac{k_B T}{3\pi(\eta_0 - \eta_{\infty})}} \zeta_2(t')\right] \,\mathrm{d}t'.
\end{equation}

Without loss of generality, we define the particle position at time $t = 0$ at which the external force is removed ($k=0$) as $x(0) = 0$. Therefore, the solution of the linear system~(\ref{eq:app1}) at subsequent times $t > 0$, averaged over a large number of realizations of the noises $\zeta_1$  and $\zeta_2$,  can be expressed in terms if the initial conditions $\dot{x}(0)$, $u(0)$ and $\dot{u}(0)$ at $t=0$ as
\begin{eqnarray}\label{eq:app2}
  	\langle x(t) \rangle & = &  \dot{x}(0) \frac{\eta_{\infty}}{\eta_0}  \tau \left[  1 - \exp\left(  -\frac{t}{\frac{\eta_{\infty}}{\eta_0}\tau}\right) \right], \nonumber\\
	\langle u(t) \rangle & =& u(0) + \dot{u}(0) \frac{\eta_{\infty}}{\eta_0}  \tau \left[  1 - \exp\left(  -\frac{t}{\frac{\eta_{\infty}}{\eta_0}\tau}\right) \right].
\end{eqnarray}

In order to obtain the initial conditions of Eq.~(\ref{eq:app2}), we take into account that at $t < 0$, the particle is dragged in a steady state by the optical trap at velocity $-v$, i.e. $\langle x(t) \rangle = -v t$. On the other hand, upon removal of the external force at $t = 0$, the ensemble average of the viscoelastic drag force $F_{drag}(t) =  \int_{-\infty}^t\Gamma(t - t')\dot{x}(t')\,\mathrm{d}t'$ in Eq.~(\ref{eq:eq2}) is zero. Therefore, after integration by parts and using the auxiliary variable $u$ defined in Eq.~(\ref{eq:appu}), we find the relation
\begin{eqnarray}\label{eq:app3}
	0 & = & \langle F_{drag}(0) \rangle,\nonumber\\
            & = & 6\pi r \left[\eta_{\infty} \dot{x}(0)  - \frac{\eta_0 - \eta_{\infty}}{\tau}u(0) \right],
\end{eqnarray}
where
\begin{eqnarray}\label{eq:app4}
	u(0) & = & \frac{1}{\tau} \int_{-\infty}^0 \langle  x(t') \rangle \exp\left( \frac{t'}{\tau}\right)\,\mathrm{d}t', \nonumber\\
            & = & v\tau.
\end{eqnarray}

Eqs.~(\ref{eq:app3}) and (\ref{eq:app4}) yield the following expressions for the initial conditions of Eq.~(\ref{eq:app2}) in terms of the initial driving velocity $v$ and the rheological quantities of the Jeffreys model
\begin{eqnarray}\label{eq:app5}
  	x(0) = 0, & & \dot{x}(0) = \frac{\eta_0 - \eta_{\infty}}{\eta_{\infty}} v, \nonumber\\
	u(0) = v \tau, & & \dot{u}(0) = -v,
\end{eqnarray}
Finally, by inserting~(\ref{eq:app5}) into Eqs.~(\ref{eq:app2}), we find the analytical expression of Eq.~(\ref{eq:eq2}) for the distance recoiled by the particle $\Delta x (t) = \langle x(t)  -  x(0) \rangle$ after removal of the imposed velocity.

\section{Particle size}\label{appb}

\begin{figure}
	\center
           \includegraphics[width=\columnwidth]{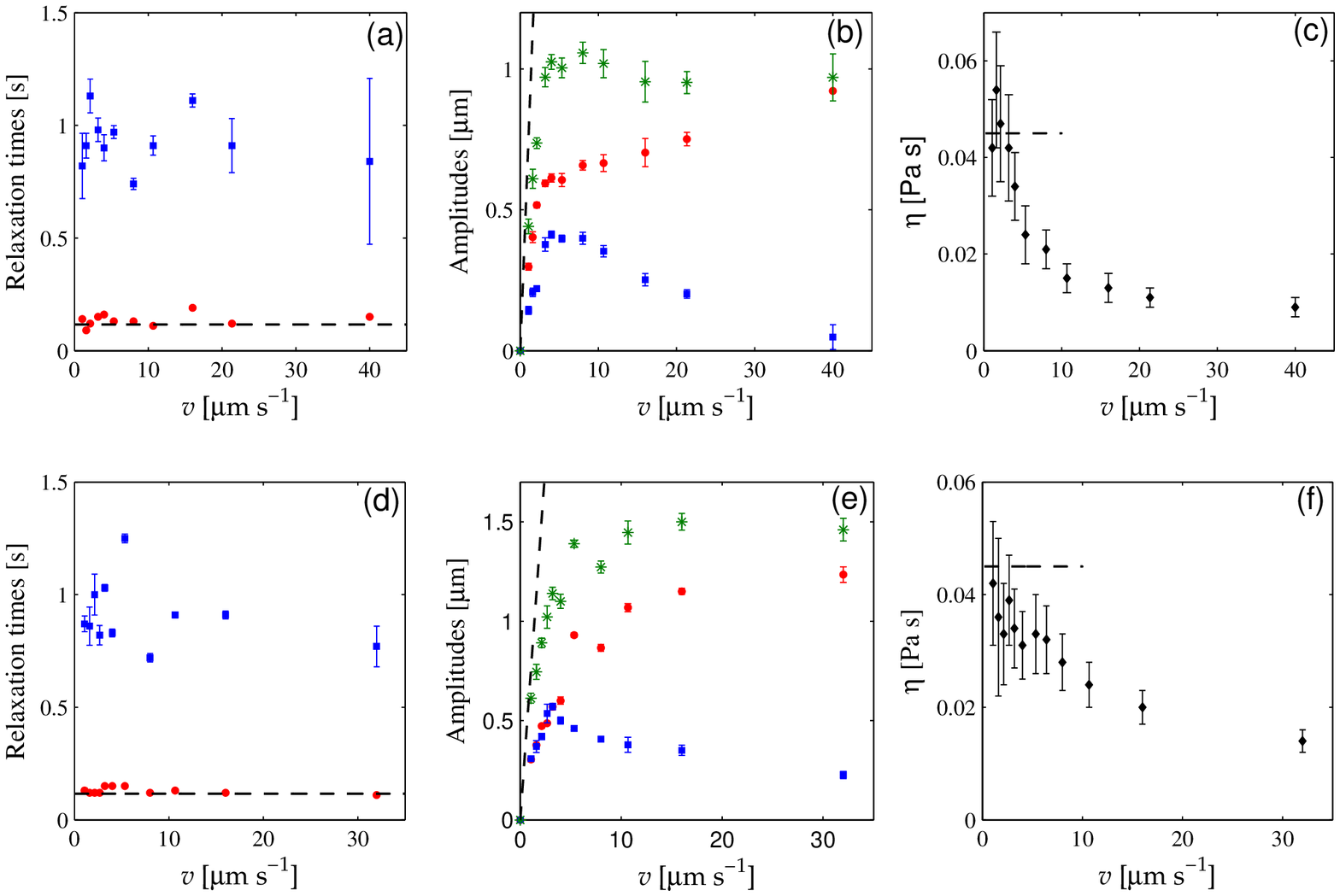}
\caption{Dependence on imposed velocity $v$ of (a) relaxation times $\tau_{\mathrm{s}}$ ($\square$), $\tau_{\mathrm{f}}$ ($\circ$), (b) amplitudes $A_{\mathrm{s}}$ ($\square$), $A_{\mathrm{f}}$ ($\circ$), $A=A_{\mathrm{s}}+A_{\mathrm{f}}$ ($*$) and (c) effective viscosity $\eta$ ($\diamond$) for a $r = 0.9\,\mu{\mathrm{m}}$ particle in a wormlike micellar solution.  Dependence on $v$ of (d) relaxation times $\tau_{\mathrm{s}}$ ($\square$), $\tau_{\mathrm{f}}$ ($\circ$), (e) amplitudes $A_{\mathrm{s}}$ ($\square$), $A_{\mathrm{f}}$ ($\circ$), and $A=A_{\mathrm{s}}+A_{\mathrm{f}}$ ($*$) and (f) effective viscosity $\eta$ ($\diamond$)for a $r = 1.3\,\mu{\mathrm{m}}$ particle in the same micellar solution. The dashed lines in Figs.~\ref{fig:figapp2}(a), \ref{fig:figapp2}(b), \ref{fig:figapp2}(d) and \ref{fig:figapp2}(e) correspond to $\tau_{\mathrm{eff}}$ and $A_{\mathrm{eff}}$ described in the main text, whereas those in Figs.~\ref{fig:figapp2}(c) and Figs.~\ref{fig:figapp2}(f) represent the zero-shear viscosity $\eta_0$ of the fluid.}
\label{fig:figapp2}
\end{figure}

In this section, we show the results for the transient dynamics of particles of different radii $r$ during the microstructural recovery of a wormlike micellar solution (CPyCl/NaSal 5 mM at $T = 303.16$~K). Figs.~\ref{fig:figapp2}(a)-(c) are the results for $r = 0.9\,\mu\mathrm{m}$ whereas those for $r =1.3\,\mu\mathrm{m}$ are plotted in Figs.~\ref{fig:figapp2}(d)-(f). We point that the dependencies of the two relaxation times $\tau_{\mathrm{f}}$ and $\tau_{\mathrm{s}}$, (Figs.~\ref{fig:figapp2}(a) and~\ref{fig:figapp2}(d)), the amplitudes  $A_{\mathrm{f}}$, $A_{\mathrm{s}}$ and  $A = A_{\mathrm{f}}+ A_{\mathrm{s}}$ (Figs.~\ref{fig:figapp2}(b) and~\ref{fig:figapp2}(e)) and the viscosities $\eta$ (Fig.~\ref{fig:figapp2}(c) and~\ref{fig:figapp2}(f)) are very similar to those shown in Fig.~\ref{fig:fig3} the main text for the same micellar solution using a particle of radius $r = 1.6\,\mu\mathrm{m}$. In particular, the values of $\tau_{\mathrm{f}}$ and $\tau_{\mathrm{s}}$ do not depend on the particle size $r$, whereas the amplitudes are independent of $r$ in the linear response regime at which $\eta = \eta_0$. The only difference is observed in the nonlinear values of the amplitudes for which thinning is previously induced at sufficiently large $v$. In such a case, Figs.~\ref{fig:figapp2}(b) and ~\ref{fig:figapp2}(e) suggest that with decreasing particle radius, the total amplitude $A$ decreases. For instance, $A \approx 1.0\,\mu\mathrm{m}$ for $r = 0.9\,\mu\mathrm{m}$,  $A \approx 1.5\,\mu\mathrm{m}$ for $r = 1.3\,\mu\mathrm{m}$, and $A \approx 1.7\,\mu\mathrm{m}$ for $r = 1.6\,\mu\mathrm{m}$. This can be explained by the fact that, at fixed $v$ and $X_0$, the smaller the particle radius $r$, the stronger the local deformation rate $v/(2r)$ and the total strain $X_0/(2r)$ of the fluid induced by the active driving, thereby resulting in a weaker nonlinear strain recovery.

\end{document}